%
%
%
\input amstex
\documentstyle{amsppt}


\topmatter
\title
Fields of Lorentz transformations on space-time
\endtitle
\rightheadtext{Fields of Lorentz transformations on space-time}

\author
Daniel Henry Gottlieb
\endauthor
\leftheadtext{Daniel Henry Gottlieb}

\thanks    
\endthanks

\address
Math. Dept., Purdue University, West Lafayette, Indiana       
\endaddress
\email
gottlieb@math.purdue.edu      
\endemail


\abstract
Fields of Lorentz transformations on a space--time $M$ are related to 
tangent bundle self isometries. In other words, a gauge transformation
with respect to the $-+++$ Minkowski metric on each fibre. Any such isometry
$L: T(M) \rightarrow T(M)$ can be expressed, at least locally, as $L = e^F$ where 
$F: T(M) \rightarrow T(M)$ is antisymmetric with respect to the metric. We find
there is a homotopy obstruction and a differential obstruction for a global
$F$. We completely study the structure of the singularity which is the 
heart of the differential obstruction and we find it is generated by
``null" $F$ which are ``orthogonal" to infinitesimal rotations $F$ with
specific eigenvalues. We find that the classical electromagnetic field
of a moving charged particle is naturally expressed using these ideas.
The methods of this paper involve complexifying the $F$ bundle maps which leads to a 
very interesting algebraic situation. We use this not only to state and
prove the singularity theorems, but to investigate the interaction of
the ``generic" and ``null" $F$, and we obtain, as a byproduct of our calculus,
a very interesting basis for the $4 \times 4$ complex matrices, and we also
observe that there are two different kinds of two dimensional complex null subspaces.
\endabstract 


\define\bG{\bold G}
\define\bC{\Bbb C}

\define\bR{\Bbb R}
\define\bZ{\Bbb Z}


\define\ds{\displaystyle}
\NoBlackBoxes

\endtopmatter
\parskip=6pt
\document 

\head 1.\ Introduction\endhead

Let $M$ be a space--time and let $T(M)$ denote its tangent bundle equipped with the Minkowski inner product $\langle \ ,\ \rangle$ of type $-+++$.
A field of Lorentz transformations refers to a vector bundle map $L\colon T(M)\to T(M)$ inducing the identity $1_M\colon M\to M$ on $M$ which preserves the inner product.
We assume that $L$ preserves orientation and the future light cones.

Now suppose $F\colon T(M)\to T(M)$ is a bundle map over the identity $1\colon M\to M$ which is skew symmetric with respect to $\langle \ ,\ \rangle$.
That is $\langle Fu,v\rangle=-\langle u,Fv\rangle$.
Then the exponential $e^F\colon T(M)\to T(M)$ is a field of Lorentz transformations, since $e^F=I+F+{F^2\over 2}+\ldots +{F^n\over n!}+\ldots$ and $\langle e^F u, e^F v\rangle=\langle e^{-F} e^F u,v\rangle=\langle e^{-F+F} u,v\rangle=\langle u,v\rangle$.
An interesting question, brought to the author's attention by Chris Phillips, 
is:\ Under what circumstances is $L=e^F$ for some $F$?

There is a homotopy obstruction.
More interestingly, there is a differential obstruction.
We show that the exponential map has a singular point at $F$ if and only if $e^F=I$ (excluding $F=0$).
The kernel of $\nabla e^F$ at the singular point is shown to consist of the ``orthogonal'' component to the two dimensional space of skew symmetric maps which commutes with $F$.

These singularities of the exponential map from $\frak s\frak o$ $(3,1)\to
SO(3,1)$ taking $F\mapsto e^F$ are very provocative physically.
We may think of a Lorentz transformation as being characterized by an orthonormal frame $\{e_0,e_1,e_2,e_3\}$.
As this frame changes under some process, it is difficult to imagine that there is not some `infinitesimal'' process $F(t)$ which yields the motion of the frame by $e^{F(t)}$.
Yet if the frame passes through its original position, there is a possibility that there is no covering infinitesimal process.

We might avoid the singularities by introducing two infinitesimal processes $F_1(t)$ and $F_2(t)$ so that $e^{F_1(t)} e^{F_2(t)}$ reproduces the motion of the frame.
But, alas, the map $\frak s\frak o(3,1)\times \frak s\frak o(3,1)\to
SO(3,1)$ also has singular points.
If we seek a nice submanifold of $\frak s\frak o(3,1)\times (\frak s\frak o(3,1)$ so that the map $(F_1,F_2)\mapsto e^{F_1}\circ e^{F_2}$ induces a proper onto map, there will still be singularities, even though the manifold itself avoids the original singularities.
Even if we consider $n$--processes, we still get singularities.

This is very closely related to the robot arm map, which takes the position of the arm and assigns it to the orientation of the end of the arm.
The existence of singularities for an $n$--linked robot arm follows from the same homotopy argument as the existence of singularities of the Generalized Robot Arm Map $\prod\limits^n \frak s\frak o(3,1)\to SO(3,1)\colon (F_1,\ldots,F_n)
\mapsto e^{F_1}\ldots e^{F_n}$.

When the robot arm approaches a singularity, the parts of the arm moves faster and faster to keep the orientation along its programmed path.
At last they drag on each other and the arm departs from the planned path.
It could remain on the path only with infinite acceleration.

What happens when a physical process can be described by a moving 4--frame $\{e_0,e_1,e_2,e_3\}$?
Does it avoid the singular points somehow; or does it pass through them, in which case, how does it react to the ``infinite accelerations'' physically?
If I could be allowed to speculate, I would say on the basis of the mathematical structure of the singularities, that their existence might give an argument of why spinning objects tend to remain spinning around a fixed or precessing axis, and when this does not obtain, there is an emission or absorption of radiation.

I base my speculation on the following observations.
Any $F\in \frak s\frak o(3,1)$, which is a skew symmetric linear transformation $F\colon \bR^{3,1}\to \bR^{3,1}$, can be classified into three classes, according to its eigenvectors.
$F$ has 1, 2 or infinitely many null eigenvector spaces (where a 
{\it null vector} $s$ means $\langle s,s\rangle=0$).
The case of infinitely many eigenvectors only occurs for the trivial case, $F=0$.
The case of one null eigenvector subspace corresponds to the null transformations, which the physicists say corresponds to the radiative case in electro--magnetism.
The generic case has two null eigenvector spaces.

The singular $F$ are infinitesimal rotations around an axis a multiple of $2\pi$ times for some observer.
The skew symmetric operators which are ``orthogonal'' to that $F$ consists of 
linear combinations
 of two null operators, one of whose eigenvectors agrees with one of the two eigenvector directions of $F$, and the other null operator's eigenvector points along the opposite eigenvector of $F$.
Now $F$ plus one of these null operators is still a rotation through an angle $2\pi n$ about an axis, but its observer is different.

Thus we see that a spinning process must maintain the same eigenvectors, hence spin axis, every $2\pi$ rotation.
And the ``directions'' having no effect on the spin are given by two null operators.

This paper depends on [Gottlieb (1998)] which may be found on my home page.
We keep the same notation.
In \S2 we will review this notation, but from a generalized complex point of view in which the main features will be stated more quickly.
In \S3, as an application of this notation, we will produce a basis for the complex $4\times 4$ matrices $M_4(\bC)$ which are Hermitian, clearly respect the decomposition of $\frak s\frak o(3,1)\otimes \bC$ into two complex quaternionic subalgebras, and in which each of the 16 matrices has a square equal to the identity.
This basis easily gives a set of multiplicative generators for the Clifford algebra $Cl(4)\simeq M_4(\bC)$.

In \S4 we will prove the singularity theorem for the exponential map $\frak s\frak o(3,1)\to  SO(3,1)$. We see that the singular points occur at
those $F$ such that $e^F = I$ and $F\neq 0$ and the image of the induced
differential on the tangent spaces is in a ``direction' parallel to $F$.
In \S5 we study adjoint mappings, finding several equations relating generic
$F$ to their two orthogonal null $F$. In \S6 we discuss the obstructions to
representing a Lorentz field  by $L = e^F$. In \S7, we show that the 
classical electromagnetic field of a moving charged particle has a very nice
description in terms of conjugation by certain exponentials of null $F$.

\head 2. Notation\endhead

We will follow the notation and conventions of [Gottlieb (1998)] as closely as possible.
But we shall outline the theory from a different point of view.
In [Gottlieb (1998)], we started with real bundle maps on the tangent space of a space--time $M$ whose tangent bundle was furnished with a Minkowskian metric
$\langle \ , \ \rangle$.
There we proceeded as geometrically as possible, defining dot products and cross products with their geometric meanings for example.
One main reason for this was to try to understand the role of choices of orientation.
As we proceeded, we added more layers of notation, creating at least three different calculi:\ Level $-2$ without choice of basis, Level $-10$ with matrices, and Level $-16$, the usual $(t,x,y,z)$ of Minkowski space.
We found, actually against our wishes, that complexifying was a powerful aid to calculation and understanding.

In this paper, we will begin with complexified Minkowski space $\bR^{3,1}\otimes \bC$, mostly use Level $-10$ notation and recover the body of [Gottlieb (1998)] by specializing to the real case and extending to fibre bundles.

Let $\bR^{3,1}$ be $\bR^4$ with an inner product $\langle \ , \ \rangle$ of type $-+++$.
The space of linear maps Hom$(\bR^{3,1},\bR^{3,1})$ is a 16--dimensional real vector space.
Let the 6--dimensional subspace of linear maps $F$ which are skew--symmetric with respect to $\langle \ ,\ \rangle$ be denoted as $\frak s\frak o(3,1)$.
Skew symmetric means $\langle Fu,v\rangle=-\langle u,Fv\rangle$.
We say that $u\in\bR^{3,1}$ is an {\it observer} if $\langle u,u\rangle=-1$ and $u$ is future pointing.

We wish to study $\bR^{3,1}\otimes \bC$.
We define the inner product $\langle \ ,\ \rangle_\bC$ on $\bR^{3,1}\otimes \bC$ by letting $\langle u,iw\rangle_\bC=\langle iv,w\rangle_\bC=i\langle v,w\rangle$ where $v,\ w\in\bR^{3,1}$.
This extends linearly to an inner product $\langle \ ,\ \rangle_\bC$ on $\bR^{3,1}\otimes \bC$.
We will usually suppress the $\bC$ and write $\langle \ ,\ \rangle$.

\demo{Remark 2.1}The usual choice of inner product on $\bR^{3,1}\otimes \bC\cong\bC^4$ is the Hermitian inner product $\langle\langle \ , \ \rangle\rangle$ which can be defined in terms of $\langle \ , \ \rangle_\bC$ by $\langle\langle v,w\rangle\rangle=\langle v,\overline w\rangle_\bC$.
The advantage of $\langle\langle \ , \ \rangle\rangle$ is that it is positive 
definite on the complex rest space of an observer, there are no null vectors ($s\not= 0$, so that $\langle\langle s,s\rangle\rangle=0$).
The advantage of $\langle \ , \ \rangle_\bC$ is that it is linear and there are null vectors!
In $\bR^{3,1}\otimes\bC$, the subspaces of null vectors are two--dimensional and they come in two distinct types.
In $\bR^{3,1}$, the null spaces are one--dimensional.
\enddemo

\demo{Scholium 2.2}
In the real case the null spaces are one--dimensional.
The propagation of light is along these one--dimensional null spaces.
One wonders what the physical meaning of the two types of null spaces is in the complex case?
After all, every one who wants to study the real physical world is forced into complexification.
God would be malicious if these null spaces had no meaning! I have a suggestion.The two different subspaces correspond to the two different helicities that a
photon can have. In other words, one subspace corresponds to a photon moving
in the same direction as its Poynting vector, and the other corresponds to a
photon moving in the opposite direction as its Poynting vector.
\enddemo

The subspace of Hom$(\bC^4)$ consisting of $F$ which are skew symmetric with respect to $\langle \ ,\ \rangle_\bC$ is denoted $\frak s\frak o(3,1)\otimes\bC$.
Thus $\langle Fv,w\rangle=-\langle v,Fw\rangle$ where $v,\ w\in \bC^4$.

The above style of notation comes from Level $-2$ of [Gottlieb (1998)].
To go to Level $-10$, we must choose an orthonormal basis $\{e_0,e_1,e_2,e_3\}$ of $\bR^{3,1}$ so that $\langle e_0,e_0\rangle=-1$ and $\langle e_i,e_j\rangle=\delta_{ij}$ otherwise.
These naturally give rise to an orthonormal basis on $\bR^{3,1}\otimes \bC\cong \bC^4$.

This orthonormal basis allows us to write linear transformations $F\colon\bC^4\to\bC^4$ as $4\times 4$ matrices, so Hom$(\bC^4)\overset\cong\to\rightarrow M_4(\bC)$, the space of $4\times 4$ complex matrices.
We see that $\frak s\frak o(3,1)\otimes \bC$ is isomorphic to the vector space consisting of all the matrices of the form
$$
\pmatrix
\vbox{
\offinterlineskip\tabskip=0pt
\halign{
\strut # &
# \hfil &
# \vrule \ \ &
\hfil # \hfil &
\hfil # \hfil &
\hfil # \hfil\cr
& 0 & & $E_1$ & $E_2$ & $E_3$\\
\omit&\multispan{5}{\hrulefill}\cr
& $E_1$ & & 0 & $B_3$ & $-B_2$\cr
& $E_2$ & & $-B_3$ & 0 & $B_1$\cr
& $E_3$ & & $B_2$& $-B_1$ & 0\cr
}}\endpmatrix
$$
We find it convenient to use the block matrix notation
$$
\pmatrix
\vbox{
\offinterlineskip\tabskip=0pt
\halign{
\strut # &
# \hfil &
# \vrule \ \ &
\hfil #\cr
& 0&&$\vec E^T$\cr
\omit&\multispan{3}{\hrulefill}\cr
&$\vec E$&&$\times B$\cr
}}\endpmatrix\text{ where }\vec E=\pmatrix E_1\\ E_2\\ E_3\endpmatrix\text{ and }\times\vec B=\pmatrix 0 & B_3 & -B_2\\ -B_3 & 0 & B_1\\ B_2 & -B_1 & 0\endpmatrix.
$$
Note that $(\times B)\vec v=\vec v\times \vec B$ if we assume that $\vec e_1\times \vec e_2=\vec e_3$.

Let us call the space of these matrices $S$ for skew symmetric.
Note they are not skew symmetric in the sense of matrices.

Now we introduce the $*$ operation on $S$.
If
$$
F=\pmatrix 0&E^T\\ E&\times B\endpmatrix,\text{ then }F^*=\pmatrix 0&-B^T\\ -B&\times E\endpmatrix.
$$

This $*$ gives rise to a mapping $F\to F^*$ on $S$ to itself which is a linear transformation and $F^{**}=-F$, so the star mapping composed with itself is $-I$ where $I\colon S\to S$ stands for the identity.

The $*$ operation is really a Level $-2$ concept.
It arises from the Hodge dual on two--forms and depends on a choice of a volume form for $\bR^{3,1}$.
Thus on level $-2$ we can define $\vec E$ and $\vec B$ for each observer $u$.
Namely $\vec E_u=Fu$ and $\vec B_u=-F^* u$ where $\vec E_u$ and $\vec B_u$ are in the rest space $T^u$ of $u$, that is $\vec E_u$ and $\vec B_u$ are orthogonal to $u$.
So $\vec E$ and $\vec B$ are freed of depending on the whole basis, just the observer.

Now define linear maps $c\colon S\to S$ and $\overline c\colon S\to S$ given by $cF\colon =F-iF^*$ and $\overline c F\colon = F+iF^*$.

\proclaim{Theorem 2.3} a) $c$ and $\overline c$ are linear maps $\frak s\frak o(3,1)\otimes \bC\to \frak s\frak o (3,1)\otimes \bC$
\item{b)}$c+\overline c=2I$
\item{c)}$c\circ \overline c=\overline c\circ c=0$
\item{d)}$c\circ c=\overline c\circ \overline c=2I$
\endproclaim

\demo{Proof}
$$
\align
c(F+G)&=(F+G)-i(F+G)^*\tag"a)"\\
&=F-iF^*+G-iG^*=cF+cG
\endalign
$$
and $c(aF)=aF-i(aF)^*=aF-i(aF^*)=a(F-iF^*)=acF$.\newline
Similarly for $\overline c$.

\noindent
b)\quad $(c+\overline c)(F)=F-iF^*+F+iF^*=2F$.

$$
\align
c\circ \overline c(F)&=c(F+iF^*)=cF+icF^*\tag"c)"\\
&=F-iF^*+i(F^*-i F^{**})\\
&=F-iF^*+iF^*+F^{**}=F+F^{**}=0.
\endalign
$$
Similarly for $\overline c\circ c=0$.

$$
\align
c\circ c(F)&=c(F-iF^*)=cF-icF^*=cF-i(F^*-iF^{**})\tag"d)"\\
&=F-iF^*-iF^*-F^{**}=2F-2iF^*=2cF.
\endalign
$$
Similarly for $\overline c\circ \overline c$.
\qed
\enddemo

The above result gives a decomposition of $S=\frak s\frak o(3,1)\otimes\bC$ into the direct sum of two subspaces, $cS\oplus\overline c S$.
In Level $-10$ notation, $c\pmatrix 0&E\\ E&\times B\endpmatrix=\pmatrix 0&A\\ A&-i(\times A)\endpmatrix$ where $\vec A=\vec E+i\vec B$, and $\overline c$ results in $\pmatrix 0&\vec A\\ \vec A&i(\times\vec A)\endpmatrix$ where $\vec A=\vec E-i\vec B$.
Note that $cS$ is the $+i$ eigenspace of $*\colon S\otimes \bC \to  S\otimes \bC$ and $\overline c S$ is the $-i$ eigenspace of $*$.
For $\pmatrix 0&\vec A\\ \vec A&\mp i\times A\endpmatrix \overset *\to\mapsto
\pmatrix 0&\pm i\vec A\\ \pm iA\vec A&\times A\endpmatrix=\pm i\pmatrix 0&A\\ A&\mp i\times A\endpmatrix$.

Now $c$ and $\overline c$ restricted to $\frak s\frak o(3,1)$ are bijections $c\colon \frak s\frak o(3,1)\overset\cong\to\longrightarrow cS$ and $\overline c\colon \frak s\frak o(3,1)\to \overline c S$.
So we have two different embedding of $\frak s\frak o(3,1)$ into $\frak s\frak o(3,1)\otimes\bC$.

Now the beautiful algebra found in [Gottlieb (1998)] follows from the following two facts.

\proclaim{Theorem 2.4}
a) Any element of $cS$ commutes with any element of $\overline cS$.

b) $F\in S$ satisfies $F^2=kI$ if and only if $F$ is in either $cS$ or $\overline c S$.
\endproclaim

\demo{Proof} Part a) is Theorem 4.8 of [Gottlieb (1998)] and part
b) is Theorem 4.5 of the same paper.

\enddemo

\proclaim{Corollary 2.5} a) $cF\overline c G=\overline c GcF$ for $F,\ G\in S$.

b) $cFcG+cGcF=kI$ for some $k$.

c) $\overline c F\overline c G + \overline c G\overline c F=\ell I$ for some $\ell$.
\endproclaim

\demo{Proof} Theorem 4.8 and Theorem 4.7 of [Gottlieb (1998)].
\enddemo

\remark{Remark 2.6} The $c\colon\frak s\frak o(3,1)\to S$ notation was used in [Gottlieb (1998)].
Now if $F$ is real, then $\overline cF=\overline{cF}$ and for eigenvalues, $\lambda_{\overline c F}=\overline\lambda_{cF}$.
This is not true in general.
This point should be understood in extending the results of \S4, \S5, \S6 to $\frak s\frak o (3,1)\otimes \bC$.
The complex conjugate should be restricted to $\overline c$.
Then most of those results should generalize.
\endremark

\proclaim{Corollary 2.7} a)\ \ $cFcG + cFcG F=2(\vec A_1\cdot \vec A_2)I$ where $cFu=\vec A_1$ and $cGu=\vec A_2$. 

We define $\langle cF,cG\rangle\colon =\langle cFu, cGu\rangle_\bC$.
This number is independent of the choice of the observer.
This is an inner product on $\frak s\frak o(3,1)\otimes\bC$.

b)\ \ $\overline c F\overline c G+\overline c G\overline c F=2(\vec A_3\cdot\vec A_4)I$ where $\vec A_3=\overline c Fu$ and $\vec F_4=\overline cGu$.

This defines another inner product $\langle F,G\rangle_- \colon =\langle \overline c Fu,\overline c Gu\rangle$ on $\frak s\frak o(3,1)\otimes \bC$.
\endproclaim

We now see from [Gottlieb (1998)] Theorem 4.3 and Corollary 4.4, that $c[F_1,F_2]u=i(c F_1 u)\times (c F_2 u)$ and $\overline c[F_1,F_2]u=-i(\overline c F_1 u)\times (\overline c F_2 u)$ and $[cF_1, cF_2]u=2i(cF_1 u)\times
(c F_1 u)$ with $[\overline c F_1,\ \overline c F_2]u=-2i(\overline c F_1 u)\times (\overline c F_2 u)$ .

In general, any $F\in \frak s\frak o(3,1)\otimes\bC$ can be written
$F={1\over 2}cF+{1\over 2}\overline c F.$
$$
\align
\ \text{ So }[F,G]u&={1\over 4}[(cF+\overline c F),\ (cG+\overline c G)]u\\
&={1\over 4}\{[cF,cG]+[\overline c F,\overline c G]\}u={1\over 4}
\{2 i cFu \times cGu+2(-i) \overline c Fu\times\overline c Gu\}\\
&={i\over 2}\{cFu \times cGu-\overline c Fu\times\overline c Gu\}.
\endalign
$$
This proves:

\proclaim{Theorem 2.8}$[F,G]u={i\over 2}\{cFu\times cGu-\overline c Fu\times\overline c Gu\}$.

\endproclaim

\head 3.\ An Interesting Basis\endhead

We will use our notation to produce an interesting basis for $M_4(\bC)$, the vector space of complex $4\times 4$ matrices.

We choose an orthogonal coordinate system $(t,x,y,z)$ for Minkowski space $\bR^{3,1}$.
Thus we are in Level $-16$ notation of [Gottlieb (1998)].
Let $e_t,\vec e_x,\vec e_y,\vec e_z$ be the unit vectors (with respect to the Minkowski metric).
We express $\vec e_x=(1,0,0)^T,\ \vec e_y=(0,1,0)^T,\ \vec e_z=(0,0,1)^T$.

Let
$$
E_x\colon =\pmatrix
\vbox{
\offinterlineskip\tabskip=0pt
\halign{
\strut # &
# \hfil &
# \vrule \ \ &
# \hfil\cr
& 0 & & $\vec e_x^T$\\
\omit&\multispan{3}{\hrulefill}\cr
& $\vec e_x$ & & $0$\cr
}}\endpmatrix
$$
and define $E_y$ and $E_z$ by replacing $\vec e_x$ by $\vec e_y$ and $\vec e_z$ respectively.

Similarly we define
$$
B_x\colon =\pmatrix
\vbox{
\offinterlineskip\tabskip=0pt
\halign{
\strut # &
# \hfil &
# \vrule \ \ &
# \hfil\cr
& 0 & & $\vec 0^T$\\
\omit&\multispan{3}{\hrulefill}\cr
& $\vec 0$ & & $\times(e_x)$\cr
}}\endpmatrix
$$
with $B_y$ and $B_z$ defined by replacing $\vec e_x$ by $\vec e_y$ and $\vec e_z$ respectively.

\proclaim{Lemma 3.1} $\{E_x, E_y, E_z, B_x, B_y, B_z\}$ form a basis for $\frak s\frak o(3,1)$ as a vector space.
\endproclaim

Now consider $cE_x,\ c E_y,\ cE_z$, and $\overline c E_x,\ \overline c E_y,\ \overline c E_z$.
Note that $\overline c E_x=\overline{c E_x}$ since $E_x$ is real.
Similarly for $y$ and $z$.

\proclaim{Lemma 3.2} $\{cE_x,\ cE_y,\ cE_z\}$ forms a basis of the complex vector space $c(\frak s\frak o(3,1))$ in $\frak s\frak o(3,1)\otimes \bC$.
Similarly $\{\overline{c E_x},\ \overline{c E_y},\ \overline{c E_z}\}$ forms a basis for the image of $\overline c$.
\endproclaim

\proclaim{Theorem 3.3} The set of sixteen matrices
$$
\matrix\format\r&\quad\r&\quad\r&\quad\r&\quad\r\\ 
I,   & cE_x c\overline{E_x},& cE_y \overline{ c E_y},& c E_z \overline{c E_z},\\
c E_x,&\overline{c E_x},&c E_y \overline{c E_z},&cE_z \overline{c E_y},\\
c E_y,&\overline{c E_y},&cE_x\overline{c E_z},&cE_z \overline{c E_x},\\
c E_z,&\overline{c E_z},&cE_x \overline{c E_y},&cE_y\overline{c E_x}
\endmatrix\tag"a)"
$$
forms a basis for $M_4(\bC)$, the vector space of $4\times 4$ complex matrices.

b) The square of each of the matrices in the basis is $I$.

c) Each matrix is Hermitian.
\endproclaim

\demo{Proof} a) Let $e_{ij}$ represent the $4\times 4$ matrix consisting of a 1 in the $i,j$ position and zeros everywhere else.
Then the sixteen $e_{ij}$ form a basis for $M_4(\bC)$.
Express the 16 matrices in a) as linear combinations of the $e_{ij}$ using the order given in a).
Then we must show that the coefficient matrix must have linearly independent rows.
Although the matrix is $16\times 16$ with 256 entries, most of the entries are zero.
In fact, the only rows which can possibly be linearly dependent form four sets.
Thus the first four rows are obviously independent of the remaining rows and similarly for the next four rows, and so on.
It is easy to show each of the four sets of four rows is linearly independent.

b) Recall that $cF^2=(\vec A\cdot \vec A)I$ where $\vec A=cF e_t=\vec E+i\vec B$.
So $c E_x c E_x=(\vec e_x\cdot \vec e_x)I=I$.
Similarly for $c E_y,\ c E_z$ and $c\overline E_x,\ c\overline E_y,\ c\overline E_z$.
Next recall that $cF\overline c G=\overline c GcF$, so $ (c\overline E_x c\overline E_y)^2=c E_x^2 c\overline E_y^2=I\cdot I=I$.
Similar arguments will work for the remaining matrices.

c) Note that the transpose of
$\pmatrix
\vbox{
\offinterlineskip\tabskip=0pt
\halign{
\strut # &
# \hfil &
# \vrule \ \ &
# \hfil\cr
& 0 & & $\vec A^T$\\
\omit&\multispan{3}{\hrulefill}\cr
& $A$ & & $\mp i(\times A)$\cr
}}\endpmatrix$ is 
$\pmatrix 
\vbox{
\offinterlineskip\tabskip=0pt
\halign{
\strut # &
# \hfil &
# \vrule \ \ &
# \hfil\cr
& 0 & & $\vec A^T$\\
\omit&\multispan{3}{\hrulefill}\cr
& $\vec A$ & & $\pm i(\times \vec A)$\cr
}}\endpmatrix$.
If $\vec A$ is a real vector, then the transpose equals the complex conjugate.
Hence 
$$
\pmatrix
\vbox{
\offinterlineskip\tabskip=0pt
\halign{
\strut # &
# \hfil &
# \vrule \ \ &
# \hfil\cr
& 0 & & $\vec A^T$\\
\omit&\multispan{3}{\hrulefill}\cr
& $\vec A$ & & $\pm i(\times \vec A)$\cr
}}\endpmatrix\text{ is Hermitian if }\vec A\text{ is real.}
$$
Thus $cE_x,\ cE_y,\ cE_z,\ \overline{c E_x},\ \overline{c E_y},\ 
\overline{c E_z}$ are all Hermitian, so is $I$.
Now $cE_x \overline{cE_y}$ is Hermitian since
$$
\align
[(cE_x)(\overline{c E_y})]^\dag&=[(\overline{cE_x})(cE_y)]^T=(cE_y)^T(\overline{c E_x})^T \\
&=(\overline{c E_y})(c E_x)=c E_x\overline{c E_y}.
\endalign
$$
The remaining matrices are Hermitian by similar arguments.
\qed
\enddemo

We can understand the multiplication properties of these matrices by the following theorem.

\proclaim{Theorem 3.4}
a) $cE_x cE_y=ic E_z=cB_z$

b) $\overline{c E_x}\ \overline{c E_y}=-i\overline{c E_z}$

c) $cE_x c E_y=-c E_y cE_x$ and $\overline{c E_x}\ \overline{c E_y}=
-\overline{c E_y}\ \overline{c E_x}     $ 

d) $cF\overline{cG}=\overline{cG}cF$
\endproclaim

\demo{Proof} c) $c E_x c E_y+c E_y c E_x=2(\vec e_x\cdot \vec e_y)I=0$ and apply complex conjugate to this equation.

a) We know that $[cE_x, cE_y] e_t=2i\vec e_x\times \vec e_y=2i\vec e_z=2i c E_z$.
But since $c E_x cE_y=-c E_y c E_x$ we have $[cE_x, cE_y]=2c E_x c E_y$.
So $(c E_x c E_y)e_t=icE_z e_t$.
Now $cFe_t$ completely determines $cF$.
So $c E_x c E_y=c E_z$.

b) Same argument or take complex conjugate of a).
\qed
\enddemo

Thus we see that $cE_x,\ cE_y,\ \overline{c E_x},\ \overline{c E_y}$ generates multiplicitivity $M_4(\bC)$, and $cE_x$ and $c E_y$ generates the Pauli
Algebra, that is, the quaternions complexified.

Now since $M_4(\bC)$ is the complex Clifford algebra $\Cal C \ell(4)$, there must be generators $\alpha_0,\alpha_1,\alpha_2,\alpha_3$ so that $\alpha_i\alpha_j+\alpha_j\alpha_i=\delta_{ij} I$.
One such set of $\alpha$'s is given by
$$
\align
\alpha_0&=c E_x\\
\alpha_1&=c E_y\\
\alpha_2&=c E_z  \overline{cE_x}\\
\alpha_3&=c E_z  \overline{cE_y}.
\endalign
$$

Of course, to obtain $\gamma_i$ such that $\gamma_i\gamma_j+\gamma_j\gamma_i=2\langle e_i,e_j\rangle I$ we let $\gamma_0=i\alpha_0,\ \gamma_i=\alpha_i$ for $i=1,2,3$ or
$$
\align
\gamma_0&=i c E_x\\
\gamma_1&= c E_y\\
\gamma_2&=c E_z\overline{c E_x}\\
\gamma_3&=c E_z\overline{c E_y}.
\endalign
$$

\head 4.\ The singularity theorem\endhead

Let $\exp\colon\frak s\frak o(3,1)\to SO(3,1)$ be the exponential map $F\mapsto e^F\colon =I+F+{F^2\over 2!}+{F^3\over 3!}+\ldots$.
Our goal is to classify the singularities of exp.

\proclaim{Singularity Theorem} The only non--regular point of exp is the identity $I\in SO(3,1)$.
Any $F \neq 0 \in\frak s\frak o(3,1)$ satisfies $e^F=I$ if and only if $\lambda_{cF}=2\pi n i$ where $n$ is a non zero integer.
The kernel of $\exp_*\colon G\mapsto \nabla_G e^F\colon = {d\over dt} (e^{F+tG})|_{t=0}$ consists of those $G\in\frak s\frak o(3,1)$ such that $\langle cF, cG\rangle=0$.
That is $\nabla_G e^F=0$ if and only if $\lambda_{cF}=2\pi ni\not= 0$ and $\langle cF,cG\rangle=0$.
\endproclaim

A general theorem for the differential of the exponential map is the following
result due to Helgason, see [Helgason (1978)], page 105, Theorem 1.7. This will help simplify our proof of the singularity
theorem.

\proclaim{Theorem (Helgason)} Let $\frak g$ be in Lie Algebra for a Lie group $\bG$.
Then $\nabla_G e^F\colon={d\over dt}(e^{F+tG})|_{t=0}=e^F\cdot (g[ad F](G))$ where $ad F\colon \frak g\to \bG$ sends $ X \longmapsto [F,X]$ and $g[\xi]$ is the power series of the function $g(\xi)=\ds{e^{-\xi}-1\over \xi}$.
\endproclaim

\proclaim{Corollary 4.1} $\nabla e^F\colon T_F \frak g\to T_F \frak g$ is singular if and only there is a non zero integer $n$ so that $2\pi i n$ is an eigenvalue of $ad(F)\colon g\to g$.
\endproclaim

\demo{Proof}$g(\xi)=0$ occurs only when $\xi\in 2\pi i(\bZ-\{0\})$.
\enddemo

\proclaim{Lemma 4.2}$2c\circ ad (F)=ad(cF)\circ c$.
\endproclaim

\demo{Proof} We must show that $2c\circ ad(F)(G)= ad (cF)(G)$ for all $G\in\frak s\frak o(3,1)$.
That is that $2c[F,G]=[cF,cG]$.
But this last is just Theorem 4.3 of [Gottlieb (1998)].
\enddemo

\proclaim{Lemma 4.3} $\exp\colon\frak s\frak o(3,1)\to SO(3,1)$ is singular at $F$ exactly when $\exp\colon S\to M_4(\bC)$ is singular at $\ds{1\over 2}cF$.
\endproclaim

\demo{Proof} Suppose $\exp\colon\frak s\frak o(3,1)\to SO(3,1)$ is singular at $F$.
Then $ad(F)\colon \frak s\frak o(3,1)\to \frak s\frak o(3,1)$ has an eigenvalue $2\pi n i\not= 0$.
Thus there is an eigenvector $cG\in \frak s\frak o(3,1)\otimes \bC$ so that $ad(F)(cG)=2\pi n i cG$.
That is $[F, cG]=2\pi n icG$.
Hence $2\pi n i cG=[F,cG]=[{cF\over 2}+{\overline{cF}\over 2}, cG]=[{cF\over 2}, cG]$.
Hence $ad({cF\over 2})\colon S\to S$ has the same eigenvalue $2\pi n i$ and hence $\nabla e^{cF\over 2}$ is singular.

Now $e^{cF}=\cosh(\lambda_{cF}) I+{\sinh(\lambda_{cF})\over \lambda_{cF}}\ cF$ according to Theorem 8.5 [Gottlieb (1998)].
Also $e^F=e^{cF\over 2} e^{\overline{cF}\over 2}$.
So $\exp\colon F \mapsto {cF\over 2} \mapsto e^{cF\over 2} \mapsto e^{\overline{cF}\over 2} e^{cF\over 2}=e^F$ is a composition of first an injective map, then an exponential map, and finally an ``absolute value'' map.
Hence the differential $\nabla e^F$ factors through the differential $\nabla e^{cF/2}$.
Hence if $cG$ is in the kernel of $\nabla e^{cF/2}$, then $G$ must be in the kernel of $\nabla e^F$.
That is $\nabla_{cG} e^{cF/2}=0$ implies $\nabla_G e^F=0$.

Now we shall show that the kernel of $\nabla e^{cF/2}$ will be 2 complex dimensional in a 3 dimensional complex space.
Hence $\nabla e^F$ has a kernel of 4 real dimensions.
But the subspace of all $G\in\frak s\frak o(3,1)$ which commutes with $F$ is two dimensional and also it cannot be in the kernel of $\nabla e^F$ by the following lemma.
Hence the kernel of $\nabla e^F$ must be 4 dimensional and so $\nabla_G e^F=0$ if and only if $\nabla_{cG} e^{cF/2}=0$.
\qed
\enddemo

\proclaim{Lemma 4.4} a) If $F$ commutes with $G\not= 0$ (in any Lie Algebra) then $\nabla_G e^F\not= 0$.

b) The two dimensional space generated by $F$ and $F^*$ commutes with $F$.
\endproclaim

\demo{Proof} a) Since $F$ commutes with $G$ we have $e^{F+tG}=e^F e^{tG}$.
Thus
$$
\align
\nabla_G e^F&={d\over dt} e^{F+tG}|_{t=0}={d\over dt} e^F e^{tG}|_{t=0}=e^F {d\over dt} e^{tG}|_{t=0}\\
&= e^F G\not= 0\text{ since }e^F \text{ is an isomorphism}.
\endalign
$$ 

b) $F^*$ commutes with $F$ by Theorem 3.2 of [Gottlieb (1998)].
\qed
\enddemo

Thus we will have proved the singularity theorem when we prove the following complex singularity theorem.

\proclaim{Theorem 4.5} a) $e^{cF}=I$ if and only if $2n\pi i=\lambda_{cF}$ and $e^{cF}=-I$ if and only if $(2n+1)\pi i=\lambda_{cF}$ for any integer $n$.

b) If $[cF, cG]=0$, then $\nabla_{cG} e^{cF}\not= 0$.

c) Assume $cF\not= 0$ and $\lambda_{cF}= n\pi i$.
Then $\nabla_{cG} e^{cF}=0$ if and only if $\langle cF, cG\rangle=0$.
\endproclaim

\demo{Proof} a) From Theorem 8.5 of [Gottlieb (1998)] we know that 
$e^{cF}=\cosh(\lambda_{cF})I+\ds{\sinh(\lambda_{cF})\over \lambda_{cF}}\ cF$.
Now $e^{cF}=I$ if and only if $\cosh(\lambda_{cF})=1$ and $\ds{\sinh(\lambda_{cF})\over \lambda_{cF}}=0$.
If $\lambda_{cF}=0$, then $e^{cF}=I+cF$, so we can exclude that case.
Hence, $\sinh \lambda_{cF}=0$ if and only if $\lambda_F=n\pi i$.
And then $\cosh(n\pi i)=\cos (n\pi)=(-1)^n$.

b) This is Lemma 4.4 a).

c) Let $A_t\colon = cF+t cG$ and let $\lambda_t\colon=\lambda_{A_t}$ be the eigenvalue of $A_t$.

Then we want to show that $\ds{d\over dt}\ \lambda_t|_{t=0}=\ds{\langle cF,cG\rangle\over \lambda_{cF}}$.

We know $A_t^2=\lambda_t^2 I=(\lambda^2_{cF}+2t \langle cF, cG\rangle + t^2 \lambda^2_{cG}) I$.
So $\ds{d\over dt} \lambda_t^2=2\langle cF,cG\rangle + 2t \lambda^2_{cG}$.

Now formally, $\ds{d\over dt}\ \lambda_t^2=2\lambda_t {d\lambda_t\over dt}$, but caution is needed here, since although $\lambda_t^2$ is a well--defined function of $t$, note that $\lambda_t$ has an ambiguity of sign and a consistent choice of sign must be made in order to make $\lambda_t$ a well--defined function of $t$.
Now if $\lambda_t$ runs over a closed loop in $\bC$ around 0, it may not be possible to define $\lambda_t$ globally.
This is no problem in our present argument, but anyone who wants to apply this formula should be careful.

So assuming $\lambda_t$ is made well--defined, we see that 
$$
{d\over dt}\lambda^2_t|_{t=0}=2\lambda_0 {d\over dt}\lambda_t|_{t=0}=2\langle cF,cG\rangle.
$$
Hence
$$
{d\lambda_t\over dt}|_{t=0}={\langle cF,cG\rangle\over \lambda_{cF}}.
\tag *
$$
Now carrying out the differention and using (*) we see that
$$
\align
\nabla_{cG} e^{cF}&={d\over dt}(\cosh \lambda_t I+{\sinh\lambda_t\over \lambda_t}(cF+tcG))|_{t=0}\\
&=\alpha I+\beta cF+\gamma cG
\endalign
$$
where
$$
\align
\alpha&=\sinh(\lambda_{cF}) {\langle cF, cG\rangle\over \lambda_{cF}}\\
\beta&={\langle cF,cG\rangle\over \lambda_{cF}^3} (\cosh (\lambda_{cF})\lambda_{cF}-\sinh(\lambda_{cF}))\\
\gamma&={\sinh\lambda_{cF}\over \lambda_{cF}}.
\endalign
$$
So $\nabla_{cG} e^{cF}=0$ if and only if $\alpha=\beta=\gamma=0$.
This follows since $cF$ and $cG$ and $I$ are linearly independent since otherwise they
would commute
Now $\gamma=0$ if and only if $\lambda_{cF}=n\pi i\not= 0$ and then $\beta=0$ if and only if $\langle cF,cG\rangle=0$.
\qed
\enddemo

\proclaim{Corollary4.6} $\nabla_{cF} e^{cF}=$
$$
\align
&=\langle cF,cG\rangle
\left\{ \left( {\sinh(\lambda_{cF})\over \lambda_{cF}}\right) I+
\left( {\cosh(\lambda_{cF})\over \lambda_{cF}^2}-{\sinh(\lambda_{cF})\over \lambda_{cF}^3}\right) cF\right\}\\
& + {\sinh(\lambda_{cF})\over \lambda_{cF}}\ cG\text{\ \  if \ \   }\lambda_{cF}\not= 0
{\text ; \ \ \ and} \\
\nabla_{cG} e^{cF}&=\langle cF,cG\rangle (I+{1\over 3} cF)+cG\text{\ \  if \ \     }\lambda_{cF}=0.
\endalign
$$
\endproclaim

\demo{Proof} The first equation comes from the above proof.
The second equation is the limit of the first equation as $cF$ is chosen so that $\lambda_{cF}\to 0$.
L'Hopital's rule is applied to the coefficients.
\enddemo

\proclaim{Corollary 4.7} $\ds{d\over dt} \lambda^2_{cF+tcG}\bigg|_{t=0}=0$ if and only if $\langle cF,cG\rangle=0$.
$\ds{d\over dt}\lambda^2_{cF+tcG}=0$ if and only if $\langle cF,cG\rangle=0$ and $cG$ is null.
\endproclaim

\head 5. Exponential equations \endhead

\proclaim{Lemma 5.1} Suppose $F$ and $G\in\frak s\frak o(3,1)$ share a null eigenvector $s$.
Then $\langle cF,cG\rangle=\lambda_{cF}\lambda_{cG}$.
In addition, one of $cG$ or $cF$ is null if and only if $cF$ and $cG$ anticommute.
\endproclaim

\demo{Proof} $(cFcG + cGcF)s=2\langle cF,cG\rangle s$ \ , hence
$$
\align
2\lambda_{cF}\lambda_{cG}&=2\langle cF,cG\rangle\\
\lambda_{cF}\lambda_{cG}&=\langle cF,cG\rangle.
\endalign
$$

Now if $cF$ is null, then $\lambda_{cF}=0$ so $\langle cF,cG\rangle=0$.
So $cF$ and $cG$ anticommute.
This argument is reversible.
\qed
\enddemo

\proclaim{Theorem 5.2} Suppose $F\in\frak s\frak o(3,1)$  and suppose $N$ is null and shares a null eigenvector $s$ corresponding to $\lambda_{cF}$.
a) Then $[cF,cN]=2\lambda_{cF} cN$.
b) If $N^{\dagger}$ is another null operator which shares the other null eigenvector of a generic $cF$, then

$$
[cN,cN^{\dagger}]={2\langle cN,cN^{\dagger}\rangle\over \lambda_{cF}}\ cF.
$$
\endproclaim

\demo{Proof} a) 
In the case where $cF$ is generic, consider $ {1 \over \lambda_ {cF}} cF$. This has eigenvalues $1$ and $-1$, and so
$ {1 \over \lambda_ {cF}} cFu$ is a unit vector. There is a choice of observer 
$u$  for which this unit vector is a real vector $\vec k$ and that
$u + \vec k$ is parallel to the real eigenvector $s$.

Now since $s$ is an eigenvector of $cN$ also, we have that $\vec i\times 
\vec j=\vec k$ where $\vec i$ and $\vec j$ are parallel to  $\vec E$ and
$\vec B$ for $cN$. That is, $\vec i$ and $\vec j$ are unit vectors and
$cNu=E(\vec i+i\vec j) $ for some constant $E$.
Now $\vec k\times (\vec i+i\vec j)=-i(\vec i+i\vec j)$.
So $cFu\times cNu=-i\lambda_{cF} cNu$ when we restore the constants to the
above equation. Now $[cF,cN]u=2i(cFu\times cNu)=2i(-i\lambda_{cF}cNu)=2\lambda_{cF}cNu$ by 
Theorem 4.3 and Corollary 4.4 of [Gottlieb (1998)]. Since the two operators
agree on an observer $u$, they must be equal, so $[cF,cN]=2\lambda_{cN}cN$.

This relation remains true when $cF$ is null since two null operators are
orthogonal if and only if they commute, and $\lambda_{cF}=0$.

b) Let $A$ and $B$ represent two null elements in $cS$ with null eigenvectors $s_+$ and $s_-$ respectively. (In order to simplify the equations, we have 
dropped
the $c$ from $cA$ and $cB$).
Let $F$ be a generic element in $cS$ with $Fs_+=\lambda_F s_+$ and $Fs_-=-\lambda_F s_-$.

Now $AB+BA=2\langle A,B\rangle I$.
Apply this equation to $s_+$ and $s_-$, making use of the facts that $As_+=0$ and $Bs_-=0$ and $\ds{Fs_{\pm} \over \lambda_F}=\pm s_\pm$.
This gives along with Theorem 5.2 a)   

$$
[A,B]s_+=2\langle A,B\rangle {Fs_+\over \lambda_F}
$$
and
$$
[A,B]s_-=2\langle A,B\rangle {Fs_-\over \lambda_F}.
$$
Adding these two equations gives
$$
[A,B](s_++s_-)=2\langle A,B\rangle {F\over \lambda_F} (s_++s_-).
$$
Since $s_+ + s_-$ is a time--like vector, this implies that
$$
[A,B]=2\langle A,B\rangle {F\over \lambda_F} \ ,
$$
 since if two skew operators
agree on the same observer they must be the same operator (see Theorem 6.9
of [Gottlieb (1998)] for a vast generalization of this last reason),
which is what was desired.
\qed
\enddemo

\proclaim{Corollary 5.3} $cF cN=\lambda_F cN$ if $cN$ is null and 
$\langle cF,cG\rangle=0$.
\endproclaim

\proclaim{Lemma 5.4} $cG cF cG=2\langle cF,cG\rangle cG-\lambda^2_{cG}cF$.
\endproclaim

\demo{Proof} Multiply $cFcG+cGcF=2\langle cF,cG\rangle I $ on the right by $cG$.
\qed
\enddemo

We will drop the $c$ in $cF$ for ease of notation 
and say instead that $F \in cS$.

\proclaim{Lemma 5.5}Let $F,G\in cS$.
Then $FG=\langle F,G\rangle I+{1\over 2} [F,G]$.
\endproclaim

\demo{Proof} $\ds{FG={1\over 2}\{F,G\}+{1\over 2}[F,G]}$.
\enddemo

\proclaim{Theorem 5.6}Let $F,G\in cS$.
Then $e^F e^G=e^D$ where
$$
\align
{\sinh \lambda_D\over \lambda_D}\ D&=(b\alpha F+a\beta G+{b\beta\over 2}\ [F,G])\text{ and } \cosh \lambda_D =
(a\alpha +b\beta \langle F,G\rangle) \ , \\
a&=\cosh \lambda_F,\ b={\sinh \lambda_F\over \lambda_F},\ \alpha=\cosh \lambda_G,\ \beta={\sinh \lambda_G\over \lambda_G}.
\endalign
$$
\endproclaim

\demo{Proof} $e^F e^G=(aI+bF)(\alpha I+\beta G)$, which follows from 
Theorem 8.5 in [Gottlieb(1998)].\newline
Expand and use the lemma above to get the equations for $D$ and $\cosh \lambda_D$
using again the expansion of $e^D$ from Theorem 8.5 to read off the equations.
\qed
\enddemo

\proclaim{Corollary 5.7} $\ds{ [e^F,e^G]={\sinh\lambda_F \sinh\lambda_G\over \lambda_F\lambda_G}\ [F,G]}$ if $F,G\in cS$.
\endproclaim

\demo{Proof} $\ds{[e^F,e^G]=[aI+bF,\ \alpha I+\beta G]=b\beta [F,G]}$.
\enddemo

\proclaim{Theorem 5.8} Let $F$ and $G$ be operators in $cS$. If $e^F=e^G\not= \pm I$, then $F=G+2\pi n\hat B$ where $\hat B$ is unit rotation, that is $\hat B$
 has eigenvalues $i$ and $-i$. In this case $F$ and $G$ commute.
\endproclaim

\demo{Proof} $0=[e^F,e^G]={\sinh \lambda_F\over \lambda_F}\ {\sinh \lambda_G\over \lambda_G}\ [F,G]$.
Hence one of $\lambda_F$ or $\lambda_G$ must equal $\pi n i$ for $n \not= 0\in\bZ$, in which case $e^F=\pm I$, say; or else $[F,G]=0$.
Hence $I=e^F e^{-G}=e^{F-G}$.
Hence $F-G=2\pi n  \hat B$ where $\hat B$ is a unit rotation.
\enddemo

\proclaim{Corollary 5.9} If $F,G\in \frak s\frak o(3,1)$ and $e^F=e^G\not= I$, then $F=G+2\pi n  \hat B$ where $\hat B$ is unit rotation in $\frak s\frak o(3,1)$.
\endproclaim

\demo{Proof} Now $e^F=e^G$ have the same null eigenvectors, so do $F$ and $G$.
Thus $[F,G]=0$.
Hence $I=e^F e^{-G}=e^{F-G}=e^{2\pi n \hat B}$.
\enddemo

\proclaim{Theorem 5.10} Let $A,B\in cS$.
Then $\lambda^2_{[A,B]}=4(\langle A,B\rangle^2-\lambda_A^2 \lambda_B^2)$.
\endproclaim

\demo{Proof} $[A,B]^2=(AB-BA)(AB-BA)$
$$
\align
&=ABAB+BABA-AB^2 A-BA^2 B\\
&=\{ABA,B\}-2\lambda_A^2 \lambda_B^2 I.\\
\text{Now }ABA&=-A^2 B+2\langle A,B\rangle A=-\lambda_A^2 B+2\langle A,B\rangle A \text{   by Lemma 5.4.}\\
\text{So }[A,B]^2 &=\{-\lambda_A^2 B+2\langle A,B\rangle A,B\}-2\lambda_A^2\lambda_B^2 I\\
&=-2\lambda_A^2 \lambda_B^2 I+4\langle A,B\rangle^2 I-2\lambda_A^2 \lambda_B^2I\\
&=4(\langle A,B\rangle^2-\lambda_A^2\lambda_B^2\rangle) I.
\endalign
$$
\qed
\enddemo

\proclaim{Corollary 5.11} If $A,B\in\frak s\frak o(3,1)$, then
$$
\lambda_{c[A,B]}^2=\langle cA,cB\rangle^2-\lambda_{cA}^2 \lambda_{cB}^2.
$$
\endproclaim

\demo{proof}
We know that $2c[F,G]=[cF,cG]$ by Theorem 4.3 of [Gottlieb (1998)]. Hence
the eigenvalue squared of $c[F,G]$ is $1/4$ times the eigenvalue squared of
$[cF,cG]$. Now apply the preceeding theorem.
\enddemo

\proclaim{Theorem 5.12}Let $A,C$ be null operators in $cS$.
Then
$$
\align
e^D&= e^A e^C=(1+\langle A,C\rangle)I+ A+C+\langle A,C\rangle \hat E\\
&=e^{A+C+\langle A,C\rangle\hat E}
\endalign
$$
where $\hat E$ is a unit boost operator orthogonal to $A$ and $B$, and
where $D$ is characterized by the vector 
$$
{\sinh \lambda_D\over \lambda_D}\ \vec D := {\sinh \lambda_D\over \lambda_D}\ Du
= \vec A + \vec C + i \vec A \times \vec C
$$

\endproclaim

\demo{Proof}$e^A e^C=(I+A)(I+C)=I+A+C+AC$
$$
\align
&=I+A+C+{1\over 2}\{A,C\}+{1\over 2}[A,C]\\
&=I+A+C+\langle A,C\rangle I+{1\over 2}(2\langle A,C\rangle \hat E)\\
&=(1+\langle A,C\rangle)I+\langle A,C\rangle \hat E+A+C.
\endalign
$$
The second to last equation follows from Theorem 5.2 b). The characterisation
of $D$ follows because
$$ [cF,cG]u = 2i \vec F \times \vec G $$
from [Gottlieb (1998) Theorem 4.3 and Corollary 4.4 as quoted in the 
paragraph after Corollary 2.7 in this paper.
\qed
\enddemo

\proclaim{Corollary 5.13} Let $A$ and $C$ be null operators in $cS$.
Then $e^A=e^C$ if and only if $A=C$.
\endproclaim

\demo{Proof}
This follows from Theorem 5.8 since $A=C +2\pi n \hat B$, and $A$ and $C$
must commute. Thus $A-C$ is null and equal to a rotation whose 
eigenvalue is $2\pi ni$. This can only happen if $n=0$.
\enddemo

\proclaim{Theorem 5.14} Let $G,\ F\in cS$.
Then
$$
\align
e^{-G}F e^G&=((\cosh \lambda_G)^2+(\sinh \lambda_G)^2)F\\
&-2\langle F,G\rangle {(\sinh \lambda_G)^2\over \lambda^2_G}\ G+
{(\sinh \lambda_G)(\cosh \lambda_G)\over \lambda_G}\ [F,G].
\endalign
$$
\endproclaim

\demo{Proof}$e^{-G}F e^G=(aI-bG)F (aI+bG)$.\newline
Expanding yields
$$
e^{-G}F e^G=a^2 F+ab FG-abGF-b^2 GFG.
$$
But
$$
GFG=-\lambda_G^2 F+2\langle F,G\rangle G.
$$
So the result follows.
\enddemo

\proclaim{Corollary 5.15} Let $F \in cS$  and $N \in cS$ be null and let $\langle F,N\rangle=0$.
Then
$$
e^{-F}N e^F=e^{-2\lambda_F}N
$$
and
$$
e^{-N} Fe^N=F+2\lambda_F N.
$$
\endproclaim

\demo{Proof}
Apply Theorem 5.2 a) and the fact that $\lambda_N = 0$ to Theorem 5.14 above
and simplify using the definition of the hyperbolic trigonometric functions.
\enddemo

\proclaim{Theorem 5.16} 
Let $F\in\frak s\frak o(3,1)$. Let $\lambda_{cF}$  be imaginary.
Then the real exponential satisfies
$$
e^F=\cos^2\left({\lambda_{F^*}\over 2}\right) I+{2\over \lambda^2_{F^*}}\sin^2\left({\lambda_{F^*}\over 2}\right) T_F+{\sin\lambda_{F^*}\over \lambda_{F^*}}\ F.
$$
If $\lambda_{cF}$ is real, then the real exponential is given by
$$
e^F=\cosh^2\left({\lambda_{F}\over 2}\right) I+{2\over \lambda^2_{F}}\sinh^2\left({\lambda_{F}\over 2}\right) T_F+{\sinh\lambda_{F}\over \lambda_{F}}\ F.
$$
If $\lambda_{cF}=0$, then the real exponential is
$$
e^F=I+F+{1 \over 2}F^2 
$$

\endproclaim

\demo{Proof} In the imaginary case we may take  $\lambda_{cF}=i\lambda_{F^*}$.
$$
\align
\text{So } e^{cF\over 2}&=\cosh \left(i{\lambda_{F^*}\over 2}\right) I+{\sinh(i{\lambda_{F^*}\over 2})\over i\lambda_{F^*}}\ cF\\
&=\cos \left({\lambda_{F^*}\over 2}\right) I+{i\sin({\lambda_{F^*}\over 2})\over i\lambda_{F^*}}\ cF\\
\text{So } e^{cF\over 2}\ e^{\overline cF\over 2}&=
\left( \cos \left({\lambda_{F^*}\over 2}\right) I+{\sin({\lambda_{F^*}\over 2})\over \lambda_{F^*}}\ cF\right) \left(\cos \left({\lambda_{F^*}\over 2}\right) I+{\sin\left({\lambda_{F^*}\over 2}\right)\over \lambda_{F^*}}\ \overline {cF}\right)\\
e^F&=\cos^2\left({\lambda_{F^*}\over 2}\right) I+{\sin\left({\lambda_{F^*}\over 2}\right)\cos \left({\lambda_{F^*}\over 2}\right)\over \lambda_{F^*}}\ (cF+\overline{cF})\\
&+ {\sin^2\left({\lambda_{F^*}\over 2}\right)\over \lambda^2_{F^*}}\ cF\overline{cF}\\
&=\cos^2\left({\lambda_{F^*}\over 2}\right) I+{2\sin\left({\lambda_{F^*}\over 2}\right)\cos\left({\lambda_{F^*}\over 2}\right)\over \lambda_{F^*}}\ F+
{2\sin^2\left({\lambda_{F^*}\over 2}\right)\over \lambda^2_{F^*}}\ T_F.
\endalign
$$
Now since $2\sin\alpha \cos\alpha=\sin (2\alpha)$, we obtain the desired result.

The real case and the null case follow similarly.
\qed
\enddemo

Note in the above theorem that $2T_F=F^2$ for null $F$. Also note
that $\left({T_F\over \lambda_T}\right)^2=I$, so ${T_F \over \lambda_T}$ is
an isometry since it is symmetric with respect to the inner product. So it
must be the exponential of something. The following results tells us what
that something is.

\proclaim{Theorem 5.17} Let $F$ be generic in $\frak s\frak o(3,1)$.
Then
$T_F=\lambda_T e^{(2n+1)\pi B}$ where $B=-{i\over \lambda_{cF}}\cdot F$ is a rotation. So $T_F/\lambda_{T_F}$ is a rotation about $180^\circ$.
\endproclaim

\demo{Proof} Note that $\lambda_{cB}=-i$.
Then
$$
e^{(2n+1)\pi B}=e^{(n+{1\over 2})\pi cB}
e^{(n+{1\over 2})\pi\overline c B}.
$$
Now $e^{(n+{1\over 2})\pi cB}=(-1)^n cB$.
Hence $e^{(2n+1)\pi B}=( (-1)^n cB)(-1)^n\overline c B)=cB\overline{cB}$.
Now $cB = -{i\over \lambda_{cF}} F$. So $cB\overline{cB} = cF \overline{cF}/
\lambda_{cF} \lambda_{\overline cF} = 2T_F/2\lambda_{T_F} = T_F/\lambda_{T_F}$.
\qed
\enddemo

\head 6. Topology and Physics\endhead

We look at the lifting problem.
Suppose we have a Gauge transformation, that is a field of Lorentz transforms, that is a bundle isometry $L$
$$
\CD
T(M)@>L>> T(M)\\
@VVpV @VVpV\\
M@>1>> M
\endCD
$$
Then we can regard $L$ as a cross--section to the associated bundle $SO(3,1)\to E@>p>> M$ where each point $e$ of $E$ is an isometry on the tangent space based on the 
point $p(e) \in M$.
Now suppose $L=e^F$ where $F\colon M\to E'$ is a cross--section where $E'\to M$ is the bundle of tangents along the fibres of $E$. It has fibres $\frak s\frak o(3,1)$.
Since $F$ can be homotopied to the zero section, it follows that its image $e^F$ is homotopic to the identity $I$.

A more concrete picture arises when we consider $M$ with trivial tangent bundle.
Then the above cross--section $L$ can be represented by a map $f\colon M\to SO(3,1)$.
Then there exists an $F$ so that $e^F=L$ only if there is a $g$ which lifts $f$ in the diagram
$$
\CD
M@>g>> \frak s\frak o(3,1)\\
@VV1V @VVexpV\\
M@>f>> SO(3,1).
\endCD
$$

This implies that $f$ is homotopic to a constant map since $\frak s\frak o(3,1)$ is contractible.
Thus it is easy to find non homotopy trivial maps which will not give rise to $e^F$.
But if $M$ is contractible, say, this homotopy obstruction does not exist.

If $I$ is in the image of $f$, then since $I$ is not regular under exp, it is possible that the maps induced on the tangent bundle are not equal,
$$
\exp_*\circ \  g_*\not= f_*\colon T_m (M)\to T_I(\frak s\frak o(3,1)),
$$
since $\exp_*\circ \   g_*$ is not onto at a critical point, whereas $f_*$ could be onto, for example.

However, there are a few interesting observations to be made:

First, $f$ may be approximated by a smooth mapping $f'$ so that $f'$ avoids $I$.
Since $M$ is 4--dimensional and $SO(3,1)$ is six dimensional, that is easy to do.

Second, if we consider $S\subset M$ a 3 dimensional space--like slice, then $f\colon S\to SO(3,1)$ maps a 3 dimensional space into a $6$ dimensional space.
We can define self intersection numbers which propagate in space--time.

Now we give an result which is similar to the singularity theorem for 
robot arms [Gottlieb (1986, 1988)]. The singularities of the robot arm can physically be seen
as infinite accelerations of the joints in the arm. The argument below follows
that of the robot arm since the map $R$ defined below is closely related to
the robot arm map.

\proclaim{Theorem 6.1} Let $R\colon \frak s\frak o(3,1)\times \ldots \times \frak s\frak o(3,1)\to SO(3,1)$ be the map $(F_1,\ldots,F_n)\mapsto e^{F_1}\ldots e^{F_n}$.
Suppose $N\subset \frak s\frak o(3,1)\times \ldots \times \frak s\frak o(3,1)$ is a submanifold so that $R\colon N\to SO(3,1)$ is onto and proper, i.e.~the 
pre-image of every compact set is compact.
Then $R$ must have a singularity.
\endproclaim

\demo{Proof} Since $R$ is proper, by a theorem of Ehresman, if $R$ has no singularities then $R$ is a fibre bundle $F\to N\to SO(3,1)$ where the fibre $F$ is a compact closed manifold.
Now $R$ is homotopy trivial since $R$ factors through $\prod\limits_n\frak s\frak o(3,1)$, a contractible space.

Hence the fibre $F$ is homotopy equivalent to $\Omega SO(3,1)\times N$.
But $\Omega SO(3,1)$ has non zero cohomology groups in infinitely many dimension
s since the connected components are homotopy equivalent of the following
series of loopspaces: $\Omega SO(3,1)\cong \Omega SO(3)\simeq \Omega (S^3)$.
Since $\Omega (S^3)$ has cohomology in infinitely many dimensions $F$ cannot be compact.
\qed
\enddemo
\head 7. A physical example\endhead

A moving particle of charge $q$ moving along a time--like path in space--time has a classical electromagnetic field given as follows.
(See [Parrott  (1987)], p.~134 for details, for example.)

Suppose the particle $p$ is situated at a point $x$ in Minkowski space--time $M$.
Suppose $p$ has 4--velocity $u\in T_x(M)$, so $\langle u,u\rangle=-1$.
Suppose $p$ undergoes an acceleration $\vec a$. Now $\vec a$ must be orthogonal
to $u$, that is $\vec a \in T_x^u$, the rest space of $p$ at $x$, or equivalently, the subspace of $T_x$ orthogonal to $u$.
Then the electro--magnetic field of $p$ is defined along the future light cone $C$ at $x$ as follows.
Any point $z$ on $C$ is given uniquely by $z=ru+r\vec w$ where $r$ is a positive number and $\vec w$ is a unit vector in the rest space of $p$ at $x$, namely $\vec w\in T_x^u$.
Now at $z$ the electric field for observer $u$ is given by $\ds{\vec E=q\left( {\vec w\over r^2}-{1\over r}\ \vec a_\perp\right)}$, whereas $\vec B=q\vec a_\perp\times \vec w$ where $\vec a_\perp\colon = \vec a-(\vec a\cdot \vec w)\vec w$ is the orthogonal part of $\vec a$ with respect to $\vec w$, Hence $\vec a_\perp\cdot \vec w=0$.

So if we let $F_{\vec a}$ stand for the skew symmetric field on $M$ due to the particle $p$, we have that $F_{\vec a}u={q\over r^2}E_{\vec w}+{q\over r} N_{\vec a}$ where $E_{\vec w}$ is the skew symmetric
operator so that $cE_{\vec w} u=\vec w$, or equivalently $E_{\vec w}$ has eigenvector along $u+\vec w$ associated with eigenvector $\lambda=1$ and $B=0$.
Hence $\lambda_{c E_w}=1$ is the eigenvalue associated with the eigenvector
$u+\vec w$.
And $N_{\vec a}$ is the null operator with $\vec E=-\vec a_\perp$ and $\vec B=\vec a_\perp\times \vec w$  eigenvector $s= a_\perp^2 u- a_\perp\times (\vec a_\perp\times \vec w)= a_\perp^2 (u+\vec w)$.
So $N_{\vec a}$ shares the eigenvector $(u+\vec w)$ with
$ E_{\vec w}$. 

\def\curl{\text{coul}}
\def\acc{N_{\vec a}}
Let $E_{\curl}=\ds{q\over r^2}\ E_{\vec w}$ represent the boost operator
corresponding to the Coulomb field of the particle at rest in its rest frame.
Then with the above notation, the following Theorem states the relationship
between the Coulomb field at rest, $\acc$, and the electro--magnetic field of the 
particle under an acceleration $\vec a$ denoted by $F_{\vec a}$.

\proclaim{Theorem 7.1} $F_{\vec a}$ is the image of $E_{\curl}$ under conjugation
by $Exp(r\acc)$. That is
$$ F_{\vec a} = e^{-rN_{\vec a}}E_{\curl} e^{rN_{\vec a}}$$

\endproclaim

\demo{Proof} Since $\acc$ and $E_{\curl}$ share an eigenvector,
$<\acc,\  E_{\curl}>=0$. Hence we may apply Corollary 5.15 to 
$cF_{\vec a}=cE_{\curl}+{q\over r} c\acc$. That gives us
$e^{-cN} cF e^{cN}=cF+2 \lambda_{cF} cN$ for null $cN$ orthogonal to $cF$.
So setting $cN$ to ${r \over 2}c \acc$ and $cF$ to $E_{\curl}$, we see that
$\lambda_{E_{\curl}}=  {q/ r^2}$ and so 
$$ e^{-{r \over 2}c \acc } cE_{\curl} e^{{r \over 2}c \acc} = cE_{\curl}+{q\over r} c\acc
= cF_{\vec a}$$

Conjugating the above equation with $e^{{r \over 2} \overline c \acc}$
yields

$$ e^{-{r}  \acc } cE_{\curl} e^{{r} \acc} = cF_{\vec a}$$
and the real part of this equation is the result to be proved.
\qed
\enddemo



\end